\documentclass[onecolumn,aps,prd,preprintnumbers,showpacs,superscriptaddress,nofootinbib,amsmath,amssymb,floats,floatfix,showkeys,notitlepage,longbibliography]{revtex4-1}
\usepackage{graphicx}	
\usepackage{amssymb}
\usepackage{dcolumn}
\usepackage{mathtools}
\usepackage{amsmath}
\usepackage{xcolor}
\usepackage{color}
\usepackage{theorem}
\usepackage{subfigure, rotating, bm, array}
\usepackage[pagebackref=false, colorlinks=true]{hyperref}
\hypersetup{linkcolor=blue, citecolor=blue,urlcolor=blue} 

\usepackage[english]{babel}           

\begin{document}

\title{Dynamical Black Holes: Apparent Horizon Versus Energy Conditions}

\author{Vitalii Vertogradov}
\affiliation{Physics department, Herzen state Pedagogical University of Russia, 48 Moika Emb., Saint-Petersburg 191186, Russia.}
\affiliation{SPB branch of SAO RAS, 65 Pulkovskoe Rd, Saint Petersburg 196140, Russia.}
\email{vdvertogradov@gmail.com}

\date{\today}

\begin{abstract}
One of the characteristics of a dynamical black hole is the apparent horizon. Initially it was assumed that this surface is spacelike. However, further studies have shown that it can be both and null and timelike. Recent studies of the black hole shadow have demonstrated the connection between energy conditions and the black hole shadow~\cite{bib:ali2024plb}. In this paper, we establish a connection between the behavior of the apparent horizon and energy conditions. We prove that the evaporation of regular black holes always leads to the formation of a horizonless object. We introduce notions of NEC horizon and prove that the apparent horizon can be null surface if and only if it coincides with the NEC horizon. We also prove that energy conditions can lead to a situation when the mass of the black hole grows and the apparent horizon decreases, which can open the way to the explanation of the information loss paradox.

\end{abstract}

\maketitle

\section{Introduction}

Einstein's general relativity predicts the existence of black holes, which were originally thought to be solely theoretical objects. However, images of supermassive black holes M87$*$~\cite{bib:m87}  and Sagittarius A$*$~\cite{bib:way} obtained by the Event Horizon Telescope Collaboration endowed these objects with real astrophysical status. Black holes are an arena for high-energy physics~\cite{bib:grib2020pocs}. They curve spacetime so strong that light can travel along circular orbits, which can be seen by a distant observer as a black spot on their sky~\cite{bib:tsupko_review, bib:ali2024podu, bib:ali2024plb}. The accretion disc surrounding a real astrophysical black hole causes it to gain mass by absorbing surrounding matter or it can lose it due to radiation processes. Therefore, the exterior spacetime needs to be time-dependent. Dynamical spacetime is crucial when examining gravitational collapse, which could lead to the formation of black holes~\cite{bib:open}  and other objects, such as naked singularity~\cite{bib:joshi_book, bib:joshi_review, bib:maharaj, bib:vertogradov2017ijmpa, bib:dey2022epjc, bib:vertogradov2022ijmpa, bib:vertogradov2024grg, bib:vertogradov2024pocs}.
Compared to static black holes, dynamical black holes are harder to analyze. First of all, the motion around dynamical black holes is more difficult to analyze because the energy of a particle is not conserved along geodesics. Nonetheless, it is still possible to either consider second order geodesic equations to gain valuable information~\cite{bib:misner, bib:yaghoub2017epjc, bib:yaghoub2018epjc, bib:vertogradov2023mpla}  or look for additional symmetries to reduce second order differential equations to first order ones~\cite{bib:germany, bib:maharaj_conformal, bib:charged_conformal, bib:yaghoub2024epjc}.
The location of horizons is another significant problem which appears in dynamical spacetimes. The notion of the event horizon, which is defined globally, is not something that can be defined in dynamical spacetime. It is possible, however,  to define it quazi-locally instead~\cite{bib:vertogradov2023mpla, bib:neldynamical, bib:nelvaidya, bib:nelsurface, bib:slowevolution, bib:quazilocal}. Marginally trapped surface, which is called an apparent horizon, is supposed to be a boundary of a black hole in dynamical spacetime. 
The apparent horizon was thought to be a closed spacelike surface. However, Hawking~\cite{bib:hawking}  demonstrated that a quantum field considered near the horizon results in the null energy condition (NEC) violation. During the Hawking radiation process, the apparent horizon appears to be time-like and shrinking. Hawking radiation has direct astrophysical consequence , specifically a reduction in the mass of black holes leads to a decreasing size of its shadow. Even as its mass increases, the shadow of a black hole can still decrease. The Theorem relating to energy conditions and black hole shadow was recently proven~\cite{bib:ali2024plb}.  According to the statement of the theorem, if the mass of the black hole grows and the NEC is satisfied, then the shadow of the black hole always increases. Once more, there are violations of NEC and a black hole shadow that is decreasing. One should note that this theorem says nothing regarding the structure of the apparent horizon. The accretion of charged particles onto a black hole is a process that can cause NEC violation. The charged Vaidya spacetime~\cite{bib:bonor} provides a good description of this accretion, despite it violates NEC near singularity~\cite{bib:charged_violation}.
The focus of this paper is on an asymptotically flat and spherically-symmetric general black hole that contains only two horizons. motivated by the fact that most of black hole models and all regular black holes~\cite{bib:dym} have only two horizons, we explore the influence of energy conditions on the structure of the apparent horizon. We introduce a notion of the NEC horizon which separates the region where null energy condition is violated from the region where it is satisfied. We formulate and prove the following assertions :
\begin{itemize}
 \item The apparent horizon is a null surface if and only if its location coincides with the NEC horizon; 
\item If the NEC horizon is either absent or located inside the black hole, the outer apparent horizon turns out to be a spacelike;
\item The outer apparent horizon is a timelike surface only if it is located inside the NEC horizon;
\item  The inner apparent horizon has a different behavior than the outer one, which is spacelike when inside the NEC horizon and timelike when outside it.
\end{itemize}
In particular, this theorem states that if a regular black hole evaporates, it always produces a horizonless object with a regular center.

The paper is organized as follows: in sec. 2 we formulate and prove the assertions listed above. We introduce the notion of NEC horizon and investigate its properties. In sec. 3 we apply our theorem to several well-known examples of charged Vaidya, regular Hayward and cosmological Vaidya spacetimes. Section IV represents discussions of obtained results and astrophysical importance.

Throughout the paper we use the geometrized system of units $G=c=1$. We adopt signature $-+++$ and dot and prime denotes partial derivatives with respect to time and radial coordinate respectively:
\begin{eqnarray}
\dot{M}&=&\frac{\partial M}{\partial v},\nonumber \\
M'&=&\frac{\partial M}{\partial r}.
\end{eqnarray}
\section{NEC versus apparent horizon}
We consider general spherically-symmetric dynamical spacetime in the form~\cite{bib:vunk}
\begin{equation} \label{eq:met}
ds^2=-\left(1-\frac{2M(v,r)}{r}\right) dv^2+2dvdr+r^2d\Omega^2,
\end{equation}
where $M(v,r)$ is the mass function of both and Eddington time $v$ and radial coordinate $r$, $d\Omega^2=d\theta^2+\sin^2\theta d\varphi^2$ is the metric on unit two-sphere. The spacetime \eqref{eq:met} is supported by energy-momentum tensor of the form
\begin{equation}
T_{ik}=\left(\rho+P\right)\left(l_in_k+l_kn_i\right)+Pg_{ik}+\mu l_il_k.
\end{equation}
Here $\rho$ and $P$ are energy density and pressure of the matter respectively, $\mu$ is energy density of energy flux and $l^i$ and $n^i$ two null vectors with property $l^in_i=-1$ and they are given by
\begin{eqnarray}
l_{i}&=&\delta^0_i,\nonumber \\
n_{i}&=&\frac{1}{2} \left (1-\frac{2M}{r} \right )\delta^0_{i}-\delta^1_{i}.
\end{eqnarray}
Physical quantities $\mu$, $\rho$ and $P$ are given by in terms of mass function as
\begin{eqnarray} \label{eq:quantity}
\mu&=&\frac{2\dot{M}}{r^2},\nonumber \\
\rho&=&\frac{2M'}{r^2},\nonumber \\
P&=&-\frac{M''}{r}.
\end{eqnarray}
Null energy condition requires
\begin{equation} \label{eq:condition}
\mu\geq 0 \rightarrow \dot{M}\geq 0.
\end{equation}
We assume that the spacetime \eqref{eq:met} describes a dynamical black hole and that it is asymptotically flat. In addition, we require that this black hole has two apparent horizons, namely the outer $r_+$ and inner $r_-$ ones. It's important to remember that a regular black hole always has only two horizons~\cite{bib:dym}\footnote{A regular black hole can also have a cosmological horizon, but in this case, the spacetime is not asymptotically flat.}. The apparent horizon equation is given via
\begin{equation} \label{eq:ah}
r_{\pm}=2M(v, r_{\pm}).
\end{equation}
Let us begin our consideration with the outer horizon. We know that, outside the black hole, one has $r\geq 2M(v,r)$. One can write $1-2M'(v,r_+)\geq 0$ where equality corresponds to an extremal black hole, i.e. when $r_-=r_+$. We assume that a black hole has two distinct horizons, and near a black hole's outer horizon, we have $1-2M'(v,r_+)>0$. Inside the outer apparent horizon, we have $r<2M(v,r)$ and we can write $1-2M'(v,r_-)\leq 0$. Again, equality corresponds to the extremal black hole, and we exclude this sign below and consider only $1-2M'(v,r_-)<0$. 

Now, we consider if these apparent horizons are timelike, spacelike or null surfaces. For this purpose we substitute \eqref{eq:ah} into \eqref{eq:met}. Note, that at the apparent horizon $f(r_{\pm})=0$. Moreover $dr_{\pm}=\dot{r}_{\pm}dv$ and $2dvdr_{\pm}$ becomes $2\dot{r}_{pm}dv^2$. Taking into account all these conditions we arrive at
\begin{equation} \label{eq:met2}
ds^2|_{r=r_{\pm}}=2\dot{r}_{\pm}dv^2+r^2_{\pm}d\Omega^2.
\end{equation}
The hypersurface $r=r_\pm$ is spacelike, timelike or null if $\dot{r}_{\pm}><=0$ respectively. Let us now calculate the sign of $\dot{r}_{pm}$
\begin{eqnarray} \label{eq:ah2}
\frac{dr_{\pm}}{dv}&=&2\frac{d}{dv}\left[M(v,r_{pm})\right]=2\dot{M}(v,r_{\pm})+2M'(v,r_{\pm})\frac{dr_{\pm}}{dv} \rightarrow\nonumber \\
\dot{r}_{\pm}&=&\frac{2\dot{M}(v,r_{\pm})}{1-2M'(v,r_{\pm})}.
\end{eqnarray}
We see that the sign of $\dot{r}_{\pm}$ depends on the apparent horizon and the sign of $\dot{M}(v,r_{\pm})$. Remember that $\dot{M}\geq 0$ is a null energy condition. Let us consider the hypersurface $r=r_{nec}$ which possesses the following condition
\begin{equation}
\dot{M}|_{r=r_{nec}}=0,
\end{equation}
i.e. this hypersurface defines the region in which NEC is violated\footnote{We assume that NEC can only be violated in the region $0\leq r<r_{nec}$. If $r_{nec}=0$, then NEC is satisfied throughout spacetime}. We will call the surface $r=r_{nec}$ as NEC horizon. When we consider the evaporation of a black hole, this assumes a violation of the NEC. As a result, we observe the following:
\begin{itemize}
\item $r_{nec}<r_-$. In this case, as can be seen from \eqref{eq:ah2}, the outer apparent horizon is a spacelike hypersurface, indicating that it grows over time. On the other hand, the inner horizon is timelike, suggesting that it shrinks over time. Therefore, the black hole's outer apparent horizon tends to infinity, while the inner horizon tends to singularity. If there is a violation of the NEC, then it intersects at $r=r_{nec}$. If NEC horizon is absent then after some time the iner apparent horizon should either disappear leaving Vaidya spacetime or asymptoticaly tends to singularity which at late times leads to Vaidya spacetime;
\item  $r_{nec}=r_-$.  In this case, the outer apparent horizon remains spacelike but the inner horizon becomes null hypersurface;
\item $r_-<r_{nec}<r_+$. If the inner horizon is inside the region of NEC violation, then the outer apparent horizon remains spacelike, but at this time the inner horizon also becomes spacelike and starts to grow. Therefore, the case where $r_{nec}$ is between horizons means that both horizons are spacelike hypersurfaces and growing;
\item $r_{nec}=r_{+}$. In this case, the inner apparent horizon remains unchanged and it remains spacelike. However, the outer horizon becomes a null surface;
\item $r_{nec}>r_+$. 
The most interesting case may correspond to the evaporation of a black hole by Hawking radiation. In this case, the inner horizon is a spacelike hypersurface and tends towards the outer apparent horizon, which, in this case, becomes a timelike hypersurface and tends towards the inner horizon. At some time $v=v_{merge}$, the apparent horizons merge, and at time $v>v_{merge}$ they disappear, leaving either a naked singularity or a regular centre, if an initial black hole was a regular one. One should note that a regular black hole contains two apparent horizons, that's why the process of black hole evaporation always leads to apparent horizon mergering. It means that when apparent horizon merge and disappear, the horizonless object with regular center is left.
\end{itemize}
Below, we give some examples of well-known black hole spacetimes where all these situations are considered. The described model can serve as a model of the formation and evaporation of a black hole with two horizons and with the assumption of asymptotic flatness.
We have explained the situation where the radius of the NEC horizon increases faster than the outer apparent horizon. However, it is important to know how this radius evolves on time because it can decrease and remain constant. One can also estimate the change of $r_{nec}$.  We know that $\dot{M}(v,r_{nec})=0$. Taking derivative with respect to $v$, one obtains
\begin{eqnarray} \label{eq:nec_speed}
\frac{d}{dv}\left(\dot{M}\right)=\ddot{M}(v,r_{nec})+\dot{M}'(v,r_{nec})\frac{dr_{nec}}{dv}=0 \rightarrow \nonumber \\
\frac{dr_{nec}}{dv}=-\frac{\ddot{M}}{\dot{M}'}.
\end{eqnarray}
One can see that the radius $r_{nec}$ can increase, decrease or be constant. 
The case of constant radius $r_{nec}$ inside the black hole is very intriguing. The inner horizon always moves towards the NEC horizon while the outer one always expands in this scenario. If the NEC is fulfilled everywhere, the inner apparent horizon tends towards singularity and spacetime becomes a Vaidya black hole asymptotically. Below, we will consider a number of examples of this process.

\section{Examples}
\subsection{Charged Vaidya spacetime}
The first example is charged Vaidya or Bonnor-Vaidya spacetime~\cite{bib:bonor}
\begin{equation} \label{eq:charged}
ds^2=-\left(1-\frac{2M(v)}{r}+\frac{Q^2(v)}{r^2}\right) dv^2+2dvdr+r^2d\Omega^2,
\end{equation}
where $M(v)$ is a mass of a black hole and $Q(v)$ its charge.
Charged Vaidya spacetime has two horizons which are given by
\begin{equation} \label{eq:ah_charged}
r_\pm=M(v)\pm \sqrt{M^2(v)-Q^2(v)}.
\end{equation}
If $Q(v) \neq 0$ then NEC is violated in charged Vaidya spacetime. The region where NEC is violated is defined as $0\leq r<r_{nec}$, where
\begin{equation} \label{eq:nec_charged}
r_{nec}=\frac{Q(v)\dot{Q}(v)}{\dot{M}(v)}.
\end{equation}
We know from the previous section that apparent horizon can be spacelike, timelike or null and it depends on the sign of $\dot{r}_\pm$, which has the form
\begin{eqnarray}
\dot{r}_{\pm}&=& \dot{M}\pm \frac{M\dot{M}-Q\dot{Q}}{\sqrt{M^2-Q^2}}=\nonumber \\
&&\dot{M}\left[1\pm \frac{M-\frac{Q\dot{Q}}{\dot{M}}}{\sqrt{M^2-Q^2}}\right]=\nonumber \\
&& \frac{\dot{M}}{\sqrt{M^2-Q^2}}\left[\pm r_{pm}\mp r_{nec}\right].
\end{eqnarray}
One can see that the apparent horizon is a null surface if $r_{\pm}=r_{nec}$. If $r_{nec}>r_+$, then the inner horizon is spacelike, and the outer is timelike, and at some time $v_{merge}$ horizons merge and then disappear. In general, when the apparent horizon contains the region of NEC violation, then it is either spacelike (outer) or timelike  (inner).
The change of $r_{nec}$ \eqref{eq:nec_speed}, in this case, is given by
\begin{equation} \label{eq:necbv}
-\frac{dr_{nec}}{dv}=\frac{\ddot{M}r^2-\dot{Q}^2r-Q\ddot{Q}r}{Q\dot{Q}}.
\end{equation}
If one considers linear mass  and charge functions 
\begin{eqnarray}
M(v)&=&\mu v,~~ \mu>0,\nonumber \\
Q(v)&=&\nu v,
\end{eqnarray}
then \eqref{eq:necbv} gives
\begin{equation}
\frac{dr_{nec}}{dv}=\frac{r_{nec}}{v}>0.
\end{equation}
It means for linear mass and charge functions the region of NEC violation grows. 

An example of constant region of NEC violation can be given by following choice of mass and charge functions
\begin{eqnarray}
M(v)&=&\mu v,~~ \mu>0,\nonumber \\
Q(v)&=&-\nu e^{-v},~~ \nu>0.
\end{eqnarray}
One should also note that if one has an electrical charge evaporation $\dot{Q}<0$, then NEC is satisfied everywhere in the spacetime. As the result, the black hole, after charge evaporation, tends to Vaidya one.
\subsection{Hayward spacetime}
In the paper~\cite{bib:hay}, Hayward offered a minimal model of a regular black hole and considered the model of its formation and evaporation. According to his model, when a black hole evaporates, then we have Minkowski spacetime. However, he only considered the evaporation of the outer horizon and didn't examine the behaviour of the inner horizon. We will show that when a black hole evaporates, the spacetime is not Minkowski because it contains mass, and the resulting object is a horizonless object with a regular centre. 
Hayward introduced his model with varying mass $M(v)$ and constant parameter $L$. We consider this spacetime with a time-dependent parameter $L$, but our results are valid for constant $L$. Thus, Hayward's spacetime is
\begin{equation} \label{eq:hay}
ds^2=-\left(1-\frac{2M(v)r^2}{r^3+2M(v)L^2(v)}\right) dv^2+2dvdr+r^2d\Omega^2.
\end{equation}
This black hole has two horizons. We don't calculate its explicit form. Instead, we use the formalism introduced in the previous section. But before doing this, we note that spacetime \eqref{eq:hay} has two horizons not for all range of functions $M(v)$ and $L(v)$:
\begin{itemize}
\item if $M(v)=\frac{3\sqrt{3}}{4}L(v)$ then the resulting object is a extremal black hole with only one horizon;
\item if $M(v)>\frac{3\sqrt{3}}{4}L(v)$ then a regular black hole has two distinct horizons;
\item if $M(v)<\frac{3\sqrt{3}}{4}L(v)$ then the central object doesn't have horizons.
\end{itemize}
The Null energy conditions demands
\begin{equation}
\dot{M}r^3-4M^2L\dot{L}\geq 0.
\end{equation}
Note, if we consider the initial model with constant $L$, then during the evaporation process $\dot{M}<0$, which means NEC is violated throughout the spacetime. From formalism developed earlier, we note that the outer horizon is timelike and shrinking, but the inner horizon is spacelike and growing. It means that at some time $v_{merge}$ when mass becomes equal to $\frac{3\sqrt{3}}{4}L$, they merge and disappear, leaving a horizonless object with non-vanishing mass, which is less than $\frac{9}{4}L$. 
Now, we return to the case when $L$ is not constant, but instead, it is a function of time $v$. We note that there is a region $0\leq r< r_{nec}$ where NEC is violated\footnote{Note, that here we don't demand $\dot{M}<0$ but we want it to be $\dot{M}>0$.}
\begin{equation}
r_{nec}=\frac{4M^2L\dot{L}}{\dot{M}}.
\end{equation}
Even if the mass of the black hole grows, we can take such $M(v)$ and $L(v)$ that at initial time $v_0$, one has
\begin{equation}
r_{null}(v_0)<r_-(v_0)<r_+(v_0),
\end{equation}
i.e. the outer horizon is spacelike, and the inner is timelike. However, if $r_{null}$ grows faster than $r_+$, then at some time $v_{violate}$ both horizons will be inside the region of NEC violation. This means that despite the growing mass of a black hole, at the time $v_{merge}$ the horizons will merge, and at the time $v>v_{merge}$ they will disappear, leaving the massive horizonless object with a regular centre. This model could serve as an example for the solution of the information loss paradox.
One also should note that if one considers $\dot{L}<0$, then NEC is satisfied everywhere in the spacetime ($\dot{M}>0$) and after vanishing, the resulting spacetime settles down to Vaidya spacetime. However, if $L=const.$ then inner horizon tends to singularity asymptotically and resulting spacetime asymptotically tends to Vaidya spacetime.
\subsection{Surrounded Vaidya Spacetime with Cosmological Fields}
The dynamical generalization of Kiselev black hole~\cite{bib:kiselev} has been found in~\cite{bib:yaghoub2017epjc} 
\begin{equation}
ds^2=-\left(1-\frac{2M(v)}{r}+\frac{N(v)}{r^{3\omega+1}}\right)dv^2+2dvdr+r^2d\Omega^2,
\end{equation}
where $\omega$ is a parameter of equation of state and $N(v)$ is cosmological parameter. The properties of apparent horizon for this spacetime have been explored in the paper~\cite{bib:yaghoub2017epjc2}.
Note that for negative values of $\omega$, the spacetime is not asymptotically flat and our method is not applicable here. So, one should consider only positive values of $\omega$. The NEC demands
\begin{equation}
\dot{M}-\frac{\dot{N}}{2r^{3\omega}}\geq 0.
\end{equation}
From this condition, one can find $r_{nec}$
\begin{equation}
r_{nec}=\left( \frac{\dot{N}}{2\dot{M}}\right)^{\frac{1}{3\omega}},~~ \omega>0.
\end{equation}
Note, that one has NEC violation only if $\dot{N}$ and  $\dot{M}$ have the same signs, i.e. $\dot{N}\dot{M}>0$.  The outer apparent horizon can be null, timelike and spacelike. However, if $\dot{N}<0$ then NEC horizon is absent and resulting spacetime tends to Vaidya spacetime.

\section{Conclusion}
The most popular asymptotically flat black hole models are equipped with two horizons. Dynamical situation states that if NEC is fulfilled throughout spacetime, the outer apparent horizon increases while the inner one decreases, resulting in a black hole with one horizon. According to the no-hair theorem, only three charges are necessary to describe a black hole: mass, electrical charge, and angular momentum. The spacetime has two horizons when more than one parameter - its mass is used to describe a black hole. If NEC is satisfied everywhere in the spacetime it means that dynamical evolution results in evaporation any charges except for mass and resulting black hole will be described with Vaidya black hole~\cite{bib:vaidya}.
As an example, charged Vaidya black hole, which evaporates its electrical charge by either accretion or charged Penrose process~\cite{bib:vertogradov2023ctp}, always satisfies NEC. This process can be observed because growing apparent horizon leads to growing black hole shadow and if initial charged black hole was nearly extremal one, then changes in black hole shadow must be significant during electrical charge evaporation process.

If NEC is violated inside the black hole then one should investigate the evolution of NEC horizon. If $r_{nec}=const.$ then the inner horizon will be frozen near the NEC horizon, i.e. it will asymptotically tends towards NEC horizon. Let us consider the situation when NEC horizon is located inside the inner horizon and $\dot{r}_{nec}>\dot{r}_+>0$. We have the following stages
\begin{itemize}
\item $r_{nec}<r_-$. In this case the inner apparent horizon is timelike and tends to singularity. The outer one is spacelike and it grows. In the observer's sky it can be seen as growing black spot;
\item $r_{nec}=r_-$. The inner horizon becomes nul but from observational point of view nothing changes because the outer apparent horizon remains a spacelike surface;
\item $r_-<r_{nec}<r_+$ Black hole spacetime has two spacelike apparent horizons and observer does not see any difference in comparison to previous two cases;
\item $r_{nec}=r_+$. The outer apparent horizon becomes null. A shadow of a black hole stops growing, and a black hole is not distinguished from a static one;
\item $r_{nec}>r_+$. The outer apparent horizon becomes timelike and starts decreasing. In this case, both horizons tend towards each other. An observer sees decreasing black spot on their sky. This situation may happen if there is an accretion of charged particles onto a black hole.
\end{itemize}
Thus, the energy condition plays a key role in the apparent horizon structure and has a direct astrophysical consequence. The accretion process has a limited impact on the evolution of black hole shadows, and it will take a considerable amount of time for observers to observe any differences in their sky. However, one should take into account the gravitational collapse because it is fast enough to be observed.
It's worth noting that this paper is solely focused on asymptotic black holes with two horizons. The issue of black holes with an undetermined number of horizons and spacetimes that are not asymptotically flat are left for future research.
\\
\textbf{acknowledgments}: Author is grateful to prof. Valerio Faraoni for scientific discussion and valuable comments.


\begin{thebibliography}{150}
\bibitem{bib:m87} The Event Horizon Telescope Collaboration, First M87
Event Horizon Telescope results.
I. The shadow of the supermassive black hole, Astrophys.
J. Lett. 875 (2019) L1
\bibitem{bib:way} Akiyama, K.~et~al. [Event Horizon Telescope Collaboration].
First Sagittarius A* Event Horizon Telescope Results. I. The Shadow of the Supermassive Black Hole in the Center of the Milky Way. \emph{Astrophys. J. Lett.} \textbf{2022}, \emph{930}, L12.
\bibitem{bib:grib2020pocs} Grib, A. A., Pavlov Y. V. (2020) Rotating
black holes as sources of high energy particles. Physics of Complex Systems, 1 (1), 40-49.DOI: 10.33910/2687-153X-2020-1-1-40-49
        \bibitem{bib:tsupko_review} V. Perlick, O.Y. Tsupko, Calculating black hole shadows: Review of analytical studies, Physics Reports, 947, 1 (2022) [arXiv:2105.07101 [gr-qc]].
\bibitem{bib:ali2024podu} V. Vertogradov, \"Ovg\"un, Analyzing the Influence of Geometrical Deformation on Photon Sphere and Shadow Radius: A New Analytical Approach -- Spherically
Symmetric Spacetimes. Phys.Dark Univ. 45, 101541 (2024)
\bibitem{bib:ali2024plb} V. Vertogradov, A. \"Ovg\"un, General Approach on Shadow Radius and Photon Spheres in Asymptotically Flat Spacetimes and the Impact of Mass-Dependent Variations.
Physics Letters B 854, 138758 (2024)
\bibitem{bib:open} J.R. Oppenheimer, H.Snyder, On Continued Gravitational Contraction. Phys. Rev. 56, 455-459 (1939).
        \bibitem{bib:joshi_book} P.S. Joshi Gravitational collapse and spacetime singularities. Cambridge University Press. 2007.p.273.
        \bibitem{bib:joshi_review} P.S. Joshi, D. Malafarina, "Recent development in gravitational collapse and spacetime singularitits". Int. J. Mod. Phys. D, 20 2641 (2011). [arXiv:1201.3660 [gr-qc]]
\bibitem{bib:maharaj}    Mkenyeleye, M.D.; Goswami, R.; Maharaj, S.D. \emph{Phys. Rev. D} \textbf{2015}, \emph{92}, 024041.
\bibitem{bib:vertogradov2017ijmpa}Vertogradov, V. The eternal naked singularity formation
in the case of gravitational collapse of generalized Vaidya spacetime. \emph{Int. J. Mod. Phys. A} \textbf{2018}, \emph{33}, 1850102. arXiv:2210.16131 [gr-qc]
\bibitem{bib:dey2022epjc} D. Dey, K. Mosani, P. Joshi, V. Vertogradov, Causal structure of singularity in non-spherical gravitational collapse. Eur. Phys. J. C 82, 431 (2022) [arXiv:2103.07190]
\bibitem{bib:vertogradov2022ijmpa} Vitalii Vertogradov. The structure of the generalized Vaidya spacetime containing the eternal naked singularity. International Journal of Modern Physics A Vol. 37, No. 28n29, 2250185 (2022) [arXiv:2209.10953.]
\bibitem{bib:vertogradov2024grg} V. Vertogradov, The generalized Vaidya spacetime with polytropic equation of state. General Relativity and Gravitation (2024) 56:59
\bibitem{bib:vertogradov2024pocs} Vertogradov, V. D. (2024) Does a primary hair have an impact on the naked singularity formation in hairy Vaidya spacetime? Physics of Complex Systems, 5 (2), 83–90
\bibitem{bib:misner}R. W. Lindquist, R. A. Schwartz, C. W. Misner, Physical Review 137, 1364 (1965).
\bibitem{bib:yaghoub2017epjc}    Heydarzade, Y.; Darabi, F. Surrounded Vaidya solution by cosmological fields. \emph{Eur. Phys. J. C} \textbf{2018}, \emph{78}, 582.
\bibitem{bib:yaghoub2018epjc}    Heydarzade, Y.; Darabi, F. Surrounded Bonnor Vaidya solution by cosmological fields. \emph{Eur. Phys. J. C} \textbf{2018}, \emph{78}, 1004.
\bibitem{bib:vertogradov2023mpla} V. Vertogradov and D. Kudryavcev, Generalized vaidya spacetime: horizons, conformal symmetries, surfacegravity and diagonalization, Modern Physics Letters A, p. 2350119, 2023, [arXiv:2212.07130 [gr-qc]]
        \bibitem{bib:germany} J. Solanki, V. Perlick, Photon sphere and shadow of a time-dependent black hole described by a Vaidya metric,  Phys. Rev. D 105, 064056, (2022)  [arXiv:2201.03274 [gr-qc]]
        \bibitem{bib:maharaj_conformal}S. Ojako, R. Goswami, S. D. Maharaj, R. Narain, Conformal symmetries in generalized Vaidya spacetimes, Class. Quantum Grav. 37 2020  055005 [arXiv:1904.08120 [gr-qc]].
        \bibitem{bib:charged_conformal}S. Koh, M. Park, A. M. Sherif, Thermodynamics with conformal Killing vector in the charged Vaidya metric,J. High Energ. Phys. 2024, 28 (2024)  [arXiv:2309.17398 [gr-qc]].
\bibitem{bib:yaghoub2024epjc} Heydarzade, Y., Vertogradov, V. Dynamical photon spheres in charged black holes and naked singularities. Eur. Phys. J. C 84, 582 (2024)
\bibitem{bib:neldynamical} Nielsen, A.B. The Spatial relation between the event horizon and trapping horizon. Class. Quant. Gravity 2010, 27, 245016, doi:10.1088/0264-9381/27/24/245016.
        \bibitem{bib:nelvaidya} A.B. Nielsen, Revisiting Vaidya horizons, Galaxies, 2, 62 (2014).
        \bibitem{bib:nelsurface} A.B. Nielsen, J.H. Yoon, Dynamical surface gravity, Class. Quant. Gravity 2008, 25, 085010 [arXiv:0711.1445 [gr-qc]].
\bibitem{bib:slowevolution} Ayon Tarafdar, Srijit Bhattacharjee, Slowly evolving horizons in Einstein gravity and beyond. [arXiv:2210.15246 [gr-qc]]
\bibitem{bib:quazilocal} Badri Krishnan, Quasi-local black hole horizons. [arXiv:1303.4635v1 [gr-qc]]
 \bibitem{bib:hawking}S. W. Hawking (1975), Particle Creation by Black Holes, Comm. Math. Phys. 43, 199.
\bibitem{bib:bonor} W. B. Bonnor and P. C. Vaidya, "`Spherically symmetric radiation of charge in Einstein-Maxwell theory"' Gen. Rel. Grav. 1, 127 (1970).
        \bibitem{bib:charged_violation} A. Ori, Charged null fluid and the weak energy condition, Class. Quantum Grav. 8 1559 (1991).
\bibitem{bib:dym} I. Dymnikova, Cosmological term as a source of mass. Class.Quant.Grav. 19 (2002) 725-740 [arXiv:gr-qc/0112052]
\bibitem{bib:vunk} A. Wang, Y. Wu, Generalized Vaidya solutions. \emph{Gen Relativ. Gravit.}  \textbf{1999}, \emph{31}, 107, [arXiv:gr-qc/9803038]
\bibitem{bib:hay} S.A. Hayward, Formation and evaporation of non-singular black holes. Phys.Rev.Lett. 96 (2006) 031103 [arXiv:gr-qc/0506126]
        \bibitem{bib:kiselev} V. V. Kiselev, Quintessence and black holes. Class. Quant. Grav. 20, 1187 (2003). [arXiv:gr-qc/0210040].
        \bibitem{bib:yaghoub2017epjc2} Y. Heydarzade, F. Darabi, Surrounded Vaidya black holes: apparent horizon properties, Eur. Phys. J. C (2018) 78:342 [arXiv:1805.01022 [gr-qc]].
        \bibitem{bib:vaidya} P.C. Vaidya,  Nonstatic solutions of Einstein's field equations for spheres of fluids radiating energy, Phys. Rev. 83, 10 (1951).
\bibitem{bib:vertogradov2023ctp} V. Vertogradov, Extraction Energy From Charged Vaidya Black Hole Via Penrose Process. Commun. Theor. Phys. 75 045404 (2023) 

\end{thebibliography}
\end{document}